\documentclass[pra,twocolumn,showpacs,superscriptaddress,floatfix]{revtex4}
\font\tenmib=cmmib10 
\font\sevenmib=cmmib7
\font\fivemib=cmmib5 

\mathchardef\Ba   = "050B  
\mathchardef\Bb   = "050C  
\mathchardef\Bg   = "050D  
\mathchardef\Bd   = "050E  
\mathchardef\Be   = "0522  
\mathchardef\Bee  = "050F  
\mathchardef\Bz   = "0510  
\mathchardef\Bh   = "0511  
\mathchardef\Bthh = "0512  
\mathchardef\Bth  = "0523  
\mathchardef\Bi   = "0513  
\mathchardef\Bk   = "0514  
\mathchardef\Bl   = "0515  
\mathchardef\Bm   = "0516  
\mathchardef\Bn   = "0517  
\mathchardef\Bx   = "0518  
\mathchardef\Bom  = "0530  
\mathchardef\Bp   = "0519  
\mathchardef\Br   = "0525  
\mathchardef\Bro  = "051A  
\mathchardef\Bs   = "051B  
\mathchardef\Bsi  = "0526  
\mathchardef\Bt   = "051C  
\mathchardef\Bu   = "051D  
\mathchardef\Bf   = "0527  
\mathchardef\Bff  = "051E  
\mathchardef\Bch  = "051F  
\mathchardef\Bps  = "0520  
\mathchardef\Bo   = "0521  
\mathchardef\Bome = "0524  
\mathchardef\BG   = "0500  
\mathchardef\BD   = "0501  
\mathchardef\BTh  = "0502  
\mathchardef\BL   = "0503  
\mathchardef\BX   = "0504  
\mathchardef\BP   = "0505  
\mathchardef\BS   = "0506  
\mathchardef\BU   = "0507  
\mathchardef\BF   = "0508  
\mathchardef\BPs  = "0509  
\mathchardef\BO   = "050A  
\mathchardef\BDpr = "0540  
\mathchardef\Bstl = "053F  

\let\a=\alpha \let\b=\beta  \let\g=\gamma  \let\d=\delta  \let\e=\varepsilon
\let\z=\zeta      \let\k=\kappa  
\let\m=\mu                  \let\r=\rho
\let\s=\sigma    \let\f=\varphi 
      
\let\G=\Gamma \let\D=\Delta   
    \let\F=\Phi

\def\CC{{\cal C}}
\def\FF{{\cal F}}
\def\II{{\cal I}}

\let\0=\noindent
\let\Dpr=\BDpr

\let\txt=\textstyle

\def\\{\hfill\break}

\def\*{\vskip3mm} 
\let\partial=\partial
\def\defi{\,{\buildrel def\over=}\,}
\def\V#1{{\bf#1}}
\def\media#1{{\langle#1\rangle}}
\def\fra#1#2{{#1\over#2}}
\def\otto{\,{\kern-1.truept\leftarrow\kern-5.truept\to\kern-1.truept}\,}
\def\tende#1{\,\vtop{\ialign{##\crcr\rightarrowfill\crcr
 \noalign{\kern-1pt\nointerlineskip} \hskip3.pt${\scriptstyle
 #1}$\hskip3.pt\crcr}}\,}

\newcommand\revtex{{R\kern-0.4mm\lower0.5mm\hbox{E}\kern-0.4mm V\kern-0.3mm%
\lower0.5mm\hbox{T}\kern-0.4mm E\kern-.3mm \lower0.5mm\hbox{X}}}
\newcommand\fancyhdr{{{F\kern-1mm\lower0.5mm\hbox{A}\kern-0.6mm N\kern-0.5mm%
\lower0.5mm\hbox{C}\kern-0.5mm Y\kern-.5mm\lower0.5mm\hbox{H}%
\kern-0.3mm\hbox{D}\kern-0.45mm\lower0.5mm\hbox{R}}}}
\newcommand\EqaligN{{{E\kern-0.3mm\lower0.5mm\hbox{Q}\kern-0.4mm A\kern-0.4mm%
\lower0.5mm\hbox{L}\kern-0.20mm I\kern-.30mm\lower0.4mm\hbox{G}%
\kern-.35mm\hbox{N}\kern-0.42mm\lower0.5mm\hbox{N}\kern-.35mm\hbox{O}}}}

\newdimen\xshift \newdimen\xwidth \newdimen\yshift \newdimen\ywidth

\def\ins#1#2#3{\vbox to0pt{\kern-#2\hbox{\kern#1 #3}\vss}\nointerlineskip}

\def\eqfig#1#2#3#4#5{
\par\xwidth=#1 \xshift=\hsize \advance\xshift
by-\xwidth \divide\xshift by 2
\yshift=#2 \divide\yshift by 2%
{\hglue\xshift \vbox to #2{\vfil
#3 \includegraphics{#4.eps}
}\hfill\raise\yshift\hbox{#5}}}

\begin{document}

\title{Irreversibility time scale}
\*

\author{\it G.Gallavotti}
\affiliation{\it I.N.F.N. Roma 1, Fisica Roma1}

\date{26 March 2006}
\begin{abstract}
Entropy creation rate is introduced for a system interacting with
thermostats ({\it i.e.} for a system subject to internal
conservative forces interacting with ``external'' thermostats via
conservative forces) and a fluctuation theorem for it is proved. As an
application a time scale is introduced, to be interpreted as the time
over which irreversibility becomes manifest in a process leading from
an initial to a final stationary state of a mechanical system in a
general nonequilibrium context. The time scale is evaluated in a few
examples, including the classical Joule-Thompson process (gas
expansion in a vacuum).
\end{abstract} 
\maketitle

\kern3mm
\def\chaos0{\bf Processes, {\it i.e.}
transformations, between equilibrium states are considered in
equilibium Thermodynamics and a few relations between macroscopic
observables are found which remain valid all along the transformation,
provided the latter can be considered as a slow sequence of successive
equilibrium states (``quasi static''). The relations are summarized by
the principles: $dU=0$ (``energy conservation'') and integrability
of the differential $(dU+pdV) T^{-1}$ (``existence of entropy''). They
can be derived from a microscopic theory in which a system is modeled
as an aggregate of molecules interacting via conservative forces.
\\ 
A corresponding theory would be desirable for stationary systems out of
equilibrium: the latter are systems subject to external
nonconservative forces and kept in stationary states by the concurrent
action of ``thermostats'' and the first question has been how to model
a thermostat.
\\ 
In fact it turned out to be quite difficult to define the temperature
out of equilibrium: here we propose that a thermostat is simply a
system of particles external to the system, and kept at constant total
kinetic energy by a force phenomenologically defined by Gauss' least
constraint principle. The constant value of the total kinetic energy
will be identified as proportional to the temperature of the
thermostat. This is quite different from what is normally called a
``isokinetic thermostat'' (often declared unphysical) which keeps the
kinetic energy of the system, rather than that of the thermostats, constant.  
\\
A further interesting proposal, in the recent literature, has been to
identify ``entropy creation rate'' as the phase space contraction
rate, {\it i.e.} as the divergence of the equations of motion;
but the notion has been criticized mainly because it is referred to
the entire set of particles including the ones in the
thermostats. This is a serious drawback because it is clear that
entropy creation, as well as any macroscopic property of the
system, should be a well defined property independent on the
thermostats size.
\\
On the other hand a general law, ``Fluctuation Theorem'', has been
derived governing fluctuations of phase space contraction rate
provided the full system (thermostats included) behaves very
chaotically (``Chaotic Hypothesis'').
\\
Here the rate of entropy generation by the system will be, differenly,
identified with the rate of entropy increase of the thermostats,
regarded as equilibrium systems with fixed temperature. It is then
shown to satisfy exactly the same asymptotic symmetry property that is
expected to hold for the total phase space contraction, {\it i.e.} the
Fluctuation Theorem, with the important feature that the property
becomes observable on a time scale substantially shorter than the time
scale necessary for observing it for all particles (for which it had
been so far proved under the Chaotic Hypothesis), including those in
the thermostats.
\\
With the entropy creation rate for the system and the temperature of
the thermostats it becomes possible to study processes 
obtained by varying the external forces and the volume and inducing an
evolution of the state of the system from an initial stationary state
to a final one. A time scale marking the waiting time necessary to
realize the irreversible nature of a process is introduced
here. Classical quasi static processes through
equilibrium states turn out to have an irreversibility time scale
$+\infty$ but the time scale makes sense also for non quasi static
processes between stationary states and is discussed here.
\\
Entropy of a stationary state is not needed nor defined, in agreement
with the doubts raised elsewhere, [18], on the possibility of defining
such a notion in stationary out of equilibrium systems.}

\def\chaos1{{\bf Consider ``processes'' between stationary
out-of-equilibrium states of a system of particles subject to external
nonconservative forces whose work is controlled by the concurrent
action of thermostats. Relevant for the phenomena are objects like
temperature, thermostats and entropy creation. Here a ``thermostat''
will simply consist of collections of particles external to the system
kept at constant total kinetic energy (identified with the thermostat
temperature) by a phenomenological force (defined by Gauss' least
constraint principle): this is quite different from what is normally
called a ``isokinetic thermostat'', keeping constant the total kinetic
energy, rather than that of the thermostats.  ``Entropy creation
rate'', (EC), is sometimes identified with ``phase space contraction
rate'', (CR), i.e. the divergence of the equations of motion; a
serious drawback is that it refers to the entire set of particles
including the ones in the thermostats. However under general
chaoticity assumptions a general law, ``Fluctuation Theorem'' (FT),
has been derived governing asymptotic fluctuations of (CR): hence a
connection between (CR) and more satisfactory definitions of (EC) is
desirable.  (EC) will be, differenly, identified with the rate of
entropy increase of the thermostats, regarded as equilibrium systems
with fixed temperature as proposed in \cite{Ja99}. It is then shown to
satisfy the very same asymptotic symmetry property expected to hold
for the total (CR), {\it i.e.} the (FT), with the extra feature that
(FT) becomes observable on a time scale substantially shorter than the
one necessary for observing it for all particles (for which it had
been so far proved) including those in the thermostats. With (EC) and
thermostats temperature it becomes possible to study processes. A time
scale marking the waiting time necessary to realize the irreversible
nature of a process is introduced here. Classical quasi static
processes through equilibrium states turn out to have an
irreversibility time scale $+\infty$ but the time scale makes sense
also for non quasi static processes between stationary states and is
discussed here.}}

\0\chaos1
\*
\centerline {\bf 1. Forces and thermostats}

In studying stationary states in nonequilibrium statistical mechanics,
\cite{Ga98,Ga04}, it is common to consider systems of particles in a
(finite) container $\CC_0$ forced by non conservative forces whose
work is controlled by {\it thermostats} consisting of other particles
moving outside $\CC_0$ in containers $\CC_a$ and interacting with the
particles of $\CC_0$ through interactions across the walls of $\CC_0$,
\cite{Ru99}.
The positions of the $N\equiv N_0$ particles in $\CC_0$ and of the $ N_a$
ones in $\CC_a$ will be denoted $\V X_a,\,a=0,\ldots,n$, and $\V
X\defi(\V X_0,\V X_1,\ldots,\V X_n)$.  Interactions 
will be described by a potential energy

\begin{eqnarray}
W(\V X)=\sum_{a=0}^{n} U_a(\V X_a) +\sum_{a=1}^n W_a(\V X_0,\V X_a)
\label{e1.1}\end{eqnarray}
{\it i.e.} thermostats particles only interact indirectly, 
via the system. All masses will be $m=1$, for
simplicity.

The particles in $\CC_0$ will also be subject to external, possibly
nonconservative, forces $\V F(\V X_0,\BF)$ depending on a few strength
parameters $\BF=(\f_1,\f_2,\ldots)$. It is convenient to imagine that
the force due to the confining potential determining the region
$\CC_0$ is included in $\V F$ so that one of the parameters is the
volume $V=|\CC_0|$.

Furthermore the thermostats particles will be also subject to
forces $\Bth$ with the property that the motions take place at constant
total kinetic energy $K_a$ that (if $k_B$ is Boltzmann's constant)
will be written as

\begin{eqnarray}K_a=\sum_{j=1}^{N_a} \fra12\, (\dot{\V X}^a_j)^2\defi
\fra32 N_a k_B T_a\defi \fra32 N_a\b_a^{-1}\label{e1.2}\end{eqnarray}
and the parameters $T_a$ will define the {\it temperatures} of the
thermostats. The exact form of the forces that have to be added in
order to insure constancy of the kinetic energies should not really
matter, within wide limits.

A choice of the thermostatting forces that has been employed in
numerical simulations has often been to define them according to
Gauss' principle of {\it least effort} for the constraints
$K_a=const$, \cite{EM90}: the simple application of the principle
yields a force on the particles of the $a$-th thermostat,
$a=1,2,\ldots$, to be, see Eq. (\ref{e1.1}),(\ref{e1.2}),

\begin{eqnarray}-\Bth_a= -
\fra{L_a-\dot U_a} {3N_a k_B T_a}\,\,\dot{\V X}_a\defi -\a_a \,\dot{\V
X}_a\label{e1.3} \end{eqnarray} 
where $L_a=-\partial_{\V X_a} W_a(\V X_0,\V X_a)\cdot \dot{\V X}_a$ is the
work done per unit time by the forces that the particles in $\CC_0$
exert on the particles in $\CC_a$. Independently of Gauss' principle
it is immediate to check that Eq.(\ref{e1.3}) implies that the motion of
the particles in $\CC_a$ conserves the kinetic energy $K_a$.


\eqfig{110pt}{90pt}{}{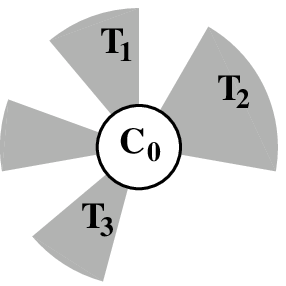}{Fig1}

\0{Caption Fig.1: The reservoirs occupy finite regions outside $C_0$,
{\it e.g.} sectors $ C_a\subset R^3$, $a=1,2\ldots$. Their particles are
constrained to have a {\it total} kinetic energy $K_a$ constant,
by suitable forces $\Bth_a$, so that the reservoirs ``temperatures''
$T_a$, see Eq.(\ref{e1.2}), are well defined.}  \*


Forces and potentials will be supposed smooth, {\it i.e.} analytic, in their
variables aside from impulsive elastic forces describing shocks,
allowed here to model shocks with the containers walls and possible
shocks between hard core particles.  In conclusion the equations of
motion of the system in its {\it phase space} $\FF$, \cite{Ga05},

\begin{eqnarray}\FF\defi(\CC_0^{N_0}\times R^{3 N_0})\times \prod_{a\ge1}
(\CC_a^{N_a}\times B_a)\equiv \FF_0\times\prod_a\FF_a\label{e1.4}\end{eqnarray}
where $B_a$ is the sphere or radius $(3 N_ak_B T_a)^{\fra12}$ in
velocity space $R^{3N_a}$, can be written as

\begin{eqnarray}
\ddot{\V X}_{0,j}=&-\partial_{\V X_{0,j}}W+\V F_j,\kern1.2cm
j=1,\ldots,N_0\cr 
\ddot{\V X}_{a,j}=&-\partial_{\V X_{a,j}}W-\Bth_{a,j},\kern1cm
j=1,\ldots,N_a\label{e1.5}
\end{eqnarray}
with $\Bth_a$ given by Eq.(\ref{e1.3}) and $\V F$ the external
> forces introduced after Eq.(\ref{e1.1}). 
Denoting $S_t(\dot{\V X},\V X)$ the configuration into which initial
data $(\dot{\V X},\V X)\defi (\dot{\V X}_a,{\V X}_a)_{a=0,\ldots,n}$
evolve in time $t$, a further condition has to be added that expresses
the physical property that thermostats are {\it efficient}. Namely for
all initial data $x=(\dot{\V X},\V X)$, {\it except possibly
for a set of data $x$ of zero volume} in $\FF$, the
limit

\begin{eqnarray}
\lim_{T\to\infty} \fra1T \int_0^T f(S_t(\dot{\V X},\V X))\,dt
=\int_\FF f(y)\m(dy)\label{e1.6}\end{eqnarray}
exists for {\it all} smooth functions $f$ and is
independent of $(\dot{\V X},\V X)$, thus defining a probability
distribution $\m$ on $\FF$. 

This of course puts restrictions on the kind of forces acting on
particles: it has to be stressed that the condition that thermostat
forces be ``efficient'' enough (impeding, for instance, an
indefinite build up of the energy) is strong but it has
to be assumed. It imposes on the potentials conditions
that are not well understood, although they seem empirically ({\it i.e.} in
simulations) verified with the simplest choices of molecular
potentials, \cite{Ga99}.

The notion of thermostat just described has evolved in the last
decades and, while it will continue to evolve, it seems to me that it
has already reached a point in which one can lay down the above
precise notion on which to base a few general considerations,
\cite{Ga05}. The class of thermostats just considered is general
enough for our purposes and is amply used in numerical simulations.  \*

The probability distribution $\m$ in Eq. (\ref{e1.6}) is called the {\it SRB
distribution} or {\it SRB statistics}. Typically one
is interested in studying time averages of observables depending only
on the system coordinates $(\dot{\V X}_0,\V X_0)$ and not on the
thermostats coordinates: in such case one needs only to know the
distribution $\m^{\CC_0}$ obtained by restricting the distribution
$\m$ to such observables: also $\m^{\CC_0}$ will be naturally called
the SRB distribution for the system.  \*

{\it Remarks}: (1) The time evolution maps $S_t,$ acting on the {\it
full} phase space $\FF$, will have the group property $S_t\cdot
S_{t'}=S_{t+t'}$ and the SRB distribution $\m$ will be invariant
under time evolution. The SRB distribution $\m$ is said to describe,
and is identified with, a {\it stationary state} of the system; it
depends on the parameters on which the forces acting on the system
depend, {\it e.g.} $|\CC_0|$ (volume) and more generally $\BF$ (strength of
the forcings), on the parameters that characterize the thermostats
like their phase space surfaces $\FF_a$, {\it i.e.} $\b_a\defi(k_B
T_a)^{-1}$, {\it etc}. The collection of SRB distributions obtained
by letting the parameters vary defines a {\it nonequilibrium
ensemble}.

(2) Finite thermostats are an idealization of the apparently more
appealing infinite thermostats in which the particles have a density
$\r_a$ and an empirical distribution which asymptotically (at
infinity) is Gibbsian with temperatures $T_a$, \cite{Ga05}: and in
such infinite thermostats the temperature should stay necessarily
constant because the local interaction between the particles in
$\CC_0$ and those in the thermostats cannot alter the velocity
distribution of the thermostat particles at large distance, at least
not if the thermostats act as physically expected.

\0(3) One could imagine replacing the forces, Eq. (\ref{e1.3}), that keep
the temperature constant, with other thermostats with which the
particles in the containers $\CC_a$ are in turn in contact: this is a
matter of where we decide to stop in our investigation of the
properties of what we call ``system''. Should we be interested in what
really happens in the containers $\CC_a$ we would have to add them to
what we call ``system'' and model their interactions with other
thermostats. In this respect the Gaussian forces that act on the
particles in the containers $\CC_a$ are a model of further thermostats
that act on the particles in $\CC_a$. If one wants to avoid completely
the forces (Gaussian as in Eq. (\ref{e1.3}) or of some other form) that fix
the temperatures of the thermostats then one would be forced to
consider infinitely extended $\CC_a$'s. But unless the infinitely
extended systems consist of non interacting particles, an important
exception, see \cite{EPR99}, not only this would lead to several
conceptual problems (like existence of solutions to the equations of
motion, phase transitions in the thermostats, nature of the
statistical properties of the motions inside the thermostats) but it
would also be possibly not needed, as in the end we are interested
only in the statistical properties of the motions in the region
$\CC_0$, {\it i.e.} on the finite system we started with.  In fact the idea
of using finite thermostats to study nonequilibrium statistical
mechanics, \cite{EM90}, has been the source of the progress in the
field in the last decades.
\*

\0{\it Remark:} Besides the mentioned relation with the ``exactly
soluble'' case of \cite{EPR99}, the above setting is very close to
\cite{Ja99} which considers what has become known as {\it transient
properties} of the fluctuations (see also \cite{GC04}): and attempts
to found on them a study of stationary states of systems like the ones
considered here.  The (CH) in this reference is present in the form of
assumptions on the correlation functions on the stationary states of
the system and thermostats. The latter works focus on thermostats
which keep their temperature constant because they are infinite
(sub)systems rather than finite isokinetic (sub)systems. The finite
thermostats, considered here, allow us to look at the problem from a
different viewpoint. It becomes possible to define entropy creation
rate, to identify it with a contribution to the phase space
contraction and to study the mechanism behind the irrelevance of the
variations of the very large amount of energy stored in the reservoirs
for what concerns the analysis of the large fluctuation relations. In
the approach of the present paper the condition of strict positivity
of the average phase space contraction plays a key role and the
restrictions on the size of the fluctuations which satisfy the (FT)
also appears in the results; and the theory of the larger fluctuations
(and their exponential tails) could be borrowed from \cite{BGGZ05}.
\*

\centerline{\bf 2. Processes}

A {\it process}, denoted $\G$, transforming an initial stationary state
$\m_{ini}\equiv \m_0$ under forcing $\BF_{ini}\equiv \BF_0$ into a final
stationary state $\m_{fin}\equiv \m_\infty$ under forcing
$\BF_{fin}\equiv \BF_\infty$ will be defined 
by a piecewise smooth function $t\to \BF(t),\, t\in[0,+\infty)$ ,
varying between $\BF(0)=\BF_0$ to
$\BF(+\infty)=\BF_\infty$. 

For intermediate times $0<t<\infty$ the time evolution $x=(\dot{\V X},\V
X) \to x(t)=S_{0,t}x$ is generated by the equations Eq. (\ref{e1.5}) with
initial state in $\FF$ and $\BF(t)$ replacing $\BF$: it is a non
autonomous equation.

The time dependence of $\BF(t)$ could for instance be due to a motion
of the container walls which changes the volume $V_t=|\CC_0|$: hence
the points $x=(\dot{\V X},\V X)$ evolve at time $t$ in a space
$\FF(t)$ which also may depend on $t$. However here no time dependence
of the thermostats temperature will be considered. It could be
introduced as in \cite{Ja99} by imagining that the potential coupling
the system and the thermostats is time dependent so that interaction
with each termostat could be switched on or off at will: such extra
freedom and generality might require extensions of what is done here.

During the process the initial state evolves into a state $\m_t$
attributing to an observable $F_t(x)$ defined on $\FF(t)$ an average
value given by

\begin{eqnarray}\media{F_t}=\int_{\FF(t)} \m_t(dx) F_t(x)\defi
\int_{\FF(0)} \m_0(dx) F_t(S_{0,t} x)\label{e2.1}\end{eqnarray}
We shall also consider the probability distribution $\m_{SRB,t}$ which
is defined as the SRB distribution of the dynamical system obtained by
``freezing''$V_t,\BF(t)$ at the value that they take at time $t$ and
imagining the time to further evolve until the stationary state
$\m_{SRB,t}$ is reached: {\it in general} $\m_t\ne \m_{SRB,t}$.  Of
course the existence of $\m_{SRB,t}$ has to be assumed {\it a priori}
and this can be implied by the general {\it Chaotic Hypothesis},
\cite{GC95}: which states that

\*
\0{\it Motions of a chaotic system, developing on
its attracting set, can be assumed to follow an evolution
with the properties of an Anosov system.}
\*

This means that in the physical problems just posed, Eq. (\ref{e1.5}) with
$\V F,\Bth$ time independent, the motions are so chaotic that the
attracting sets on which their long time motion takes place can be
regarded as smooth surfaces on which motion is highly unstable. More
precisely

\*

\0(i) around every point of $\FF$ three planes $M_s(x),$
$M_u(x),$ $M_m(x)$ can be identified which vary continuously with $x$,
which are {\it covariant} ({\it i.e.} the planes at a point $x$ are are
mapped, by the evolution flow $S_t$, into the corresponding planes
around $S_t x$) and
\\ (ii) the planes, called {\it stable, unstable and marginal}, with
respective positive dimensions $d_s>0,d_u>0$ and $d_m=1$ adding up to
the dimension of $\FF$, have the property that infinitesimal lengths
on the stable plane and on the unstable plane of any point contract at
exponential rate as time proceeds towards the future or towards the
past, respectively. The length along the marginal direction neither
contracts nor expands ({\it i.e.} it varies relative to the initial value staying
bounded away from $0$ and $\infty$): its tangent vector is parallel to
the velocity in phase space. In cases in which time evolution is
discrete, and determined by a map $S$, the marginal direction is
missing.
\\ (iii) over a time $t$, positive for the stable plane and negative
for the unstable plane, the lengths contraction is exponential, {\it i.e.}
lengths contract by a factor uniformly bounded by $C e^{-\k |t|}$ with
$C,\k>0$.
\\ (iv) there is a dense trajectory.

\kern2pt 
The Ergodic Hypothesis provides us with an expression for the equilibrium
averages (as integrals over the normalized Liouville distribution on
the energy surface): likewise the Chaotic Hypothesis provides us with
the {\it existence} and a {\it formal expression} for the averages
({\it i.e.} for the SRB distribution), \cite{GBG04}.

Of course the hypothesis is only a ``beginning'' and one has to
learn how to extract information from it, as it was the case with 
the Liouville distribution once the Ergodic Hypothesis
guaranteed it as an appropriate distribution for the study of
the statistics of motions in equilibrium, \cite{BGGZ05}.
\*

\centerline{\bf 3. Heat generation}
\*

The work $L_a$ in Eq. (\ref{e1.3}) will be interpreted as {\it heat} $\dot
Q_a$ ceded, per unit time, by the particles in $\CC_0$ to the $a$-th
thermostat (because the ``temperature'' of $\CC_a$ remains constant).
The entropy creation rate due to heat exchanges between the system and
the thermostats can, therefore, be naturally defined as

\begin{eqnarray}\s^0(\dot{\V X},\V X)\defi\sum_{a=1}^{N_a}
\fra{\dot Q_a}{k_B T_a}\label{e3.1}
\end{eqnarray}

\0{\it Remarks:} (1) It is natural to define Eq. (\ref{e3.1}) as
``heat-exchanges entropy creation'' rate because it is the amount of
work per unit time that the system performs on the thermostats which
nevertheless keep their temperature (identified with the kinetic
energy) constant.

\0(2) The above definition of ``heat exchanges entropy creation'' {\it
does not} require any definition of entropy itself for the system: it
only requires the notion of entropy variation of a reservoir
considered as a system able to absorb work without changing it into
mechanical work nor into its temperature variations. The latter is the
``classical'' definition of entropy variation of a reservoir.

\0(3) If the thermostats temperatures are equal, {\it i.e.}
$K_a\sim\fra32{N_a}\b^{-1}$ and $\BF=\V0$, or $\BF$ is conservative so
that its potential can be imagined merged in $U_0$, it is possible to
check that the SRB distribution, if existing, is necessarily

\begin{eqnarray}
\m_{SRB}(dx)=&
Z^{-1}e^{-\b (K_0+U_0+\sum_{a>0} U_a+\sum_{a>0} W_a)}\cdot\cr
&\cdot \prod_{a>0}\d({\txt K_a-\frac{3N_a-1}{2\b}})\cdot d\V V\,d\V X
\label{e3.2}
\end{eqnarray}
where $dx=d\V V\,d\V X\,=$ phase space volume element, $K_a$ are the
kinetic energies of the various subsystems, $a=0,1,\ldots$ and $Z$ is
a normalization factor. Hence this is a Gibbs distribution in one of
the equivalent, if unusual, {\it equilibrium} ensembles at temperature
$(k_B\b)^{-1}$ and densities $\r_a=\frac{N_a}{|\CC_a|},\ a=0,\ldots,n$.
This is an important consistency check, following (and extending)
\cite{EM90}.  \*

\centerline{\bf 4, Irreversibility degree}

It is one of the basic tenets in Thermodynamics that all (nontrivial)
processes between equilibrium states are ``irreversible'': only
idealized (strictly speaking nonexistent) ``quasi static'' processes
can be reversible.  The question that is addressed in the following is
whether irreversibility can be made a quantitative notion at least
in models based on microscopic evolution, Eq. (\ref{e1.5}).

Write Eq. (\ref{e1.5}) as $\ddot{\V X}=\BX(\dot{\V X},\V X)=\BX(x)$ with
$x=(x_1,$ $\ldots,\dot x_1, \ldots,\dot x_{3 N_{tot}}),\, N_{tot}=\sum_a
N_a$. Given a process $\G$ let

\kern-3mm
\begin{eqnarray}\s^\G_t(x)\defi-{\rm divergence}(x)\equiv -\sum_{i=1}^{3 N_{tot}}
\Dpr_{\dot {x}_{i}} \BX_{i}(x)\label{e4.1}\end{eqnarray}
which has the meaning of rate at which the phase volume around $x$
contracts because of the forces action.

Phase space volume can also change because new regions become
accessible (or inaccessible) so that the total phase space contraction
rate, denoted $\s_{tot,t}$, in general will be different from
$\s^\G_t$.

It is reasonable to suppose, and in some cases it can even be proved
({\it e.g.} in the example considered in remark (3) of Sect.5),
that at every time $t$ the configuration $S_{0,t}x$ is a ``typical''
configuration of the ``frozen'' system if the initial $x$ was typical
for the initial distribution $\m_0$: {\it i.e.} it will be a point in
$\FF(t)=V_t^N\times R^{N}\times \prod \FF_a$ whose statistics under
forces imagined frozen at their values at time $t$ will be
$\m_{SRB,t}$, see comments following Eq.(\ref{e2.1}). Since we consider as
accessible the phase space occupied by the attractor of a typical
phase space point, the volume variation contributes an extra
$\s^v_t(x)$ to the phase space variation via the rate at which the
volume $|\FF_t|$ contracts, namely

\kern-3mm
\begin{eqnarray}\s^v_t(x)=-\frac1{|\FF_t|}\frac{d \,|\FF_t|}{dt}=
-N\frac{\dot V_t}{V_t}\label{e4.2}\end{eqnarray}
which does not depend on $x$ as it is a property of the phase space
available to (any typical) $x$.  A calculation of the divergence
$-\s^\G$ of Eq. (\ref{e1.5}), see also Eq. (\ref{e1.3}), shows that
(keeping properly into account that the momentum part of phase space 
is a sphere)

\kern-3mm
\begin{eqnarray}\txt\s^\G(\dot{\V X},\V X)={\mathop\sum\limits_{a>0}}
{\txt\frac{3N_a-1}{3 N_a}}
  \frac{\dot Q_a-\dot U_a}{k_B T_a}
\simeq {\mathop\sum\limits_{a>0}} \frac{\dot Q_a}{k_B T_a}-\dot U
\label{e4.3}\end{eqnarray}
if $U\defi\sum_a \frac{(3N_a-1)}{3 N_a} \frac{U_a}{k_B T_a}$; in the
last step $O(N_a^{-1})$ has been neglected (for simplicity and for
consistence with macroscopic thermodynamics where $N_a$ are extremely
large: it would not be difficult to keep track of it,
however). Therefore the total phase space contraction per unit time
can be expressed as, see Eq. (\ref{e1.3}),Eq. (\ref{e3.1}),

\kern-3mm
\begin{eqnarray}\s_{tot}(\dot{\V X},\V X)= \sum_a \frac{\dot Q_a}{k_B
  T_a}-N\frac{\dot V_t}{V_t}
-\dot U\label{e4.4}
\end{eqnarray} 
{\it i.e.} there is a simple and direct relation between phase space
contraction and entropy creation rate, \cite{Ga05}.  Eq. (\ref{e4.3})
shows that their difference is a ``total time derivative''.  

In studying stationary states $N\dot V_t/V_t=0$, and the interpretation
of $\dot U$ is of ``reversible'' heat exchange
between  system and thermostats.  In this case, in some
respects, the difference $\dot U$ can be ignored. For instance in the
study of the fluctuations of the average over time of entropy creation
rate in a {\it stationary state} the term $\dot U$ gives no
contribution, or it affects only the very large fluctuations,
\cite{CV03a,BGGZ05} if the containers $\CC_a$ are very large (or if
the forces between their particles can be unbounded, which we are
excluding here for simplicity, \cite{BGGZ05}).

Even in the case of processes the quantity $\dot U$ has to oscillate
in time with $0$ average on any interval of time $(t,\infty)$
if the system starts and ends in a stationary state.

For the above reasons we define the {\it entropy creation rate} in a process
to be Eq. (\ref{e4.4}) {\it without} the $\dot U$ term:

\kern-3mm
\begin{eqnarray}\e(\dot{\V X},\V X)= \sum_a \frac{\dot Q_a}{k_B
  T_a}-N\frac{\dot V_t}{V_t}\label{e4.5}\end{eqnarray}
It is interesting, and necessary, to remark that in a stationary state
the time averages of $\e$, denoted $\e_+$, and of $\sum_a \frac{\dot
Q_a}{k_B T_a}$, denoted $\sum_a \frac{\dot Q_{a+}}{k_B T_a}$, coincide
because $N\dot V_t/V_t=0$, as $V_t=const$, and $\dot U$ has zero time
average being a total derivative. On the other hand under very general
assumptions, much weaker than the Chaotic Hypothesis, the time average
of the phase space contraction rate is $\ge0$, \cite{Ru96,Ru97}, so
that in a stationary state:

\kern-5mm
\begin{eqnarray}\e_+\equiv \sum_a \frac{\dot Q_{a+}}{k_B T_a}\ge0
\label{e4.6}\end{eqnarray}
which is a consistency property that has to be required for any
proposal of definition of entropy creation rate.
\*

\0{\it Remarks:} (1) In stationary states the above models are a
realization of {\it Carnot machines}: the machine being the system in
$\CC_0$ on which external forces $\V F$ work leaving the system in the
same stationary state (a special ``cycle'') but achieving a transfer
of heat between the various thermostats (in agreement with the second
law, see Eq. (\ref{e3.1}),Eq. (\ref{e4.6}), only if $\e_{+}\ge0$).  
\\
(2) In a stationary state $\s_{tot}=\sum_a\frac{\dot Q_a}{k_B T_a}-\dot
U$ will satisfy, under the Chaotic Hypothesis of Sec.2, a
fluctuation relation, see \cite{GC95,Ga95b}, since the dynamics
Eq. (\ref{e1.5}) is {\it time reversible}: the large fluctuations of the quantity
$p=\frac1{\e_+}\frac1T\int_0^T(\sum_{a>0}\frac{\dot Q_a}{k_B T_a}+\dot U)\,dt$
will be controlled by a {\it rate function} $\z(p)$ satisfying
$\z(-p)=\z(p)-p\e_+$, see \cite{Ga95b,BGGZ05}. However the term $\dot
U$ will not contribute to the relation because $U$ is bounded,
\cite{BGGZ05}. Therefore the fluctuations of
$p=\frac1{\e_+}\frac1T\int_0^T\sum_{a>0} \frac{\dot Q_a}{k_B T_a}\,dt$ will
have exactly the {\it same} rate function.
\\
(3) Note, however, that $U$ can be very large (being of the order of
$\sum_{a>0} N_a$) if the thermostats are large and its contribution to
the averages of $\s_{tot}$ tends to $0$ as slowly as
$T^{-1} O(\sum_{a>0} N_a)$. Hence the time necessary to see the
fluctuation relation satisfied with prefixed accuracy when $p$ is
defined as the average of $\s_{tot}$ may be enormously larger than the
time necessary to see the fluctuation relation satisfied with $p$
defined as the average of $\sum_a\frac{\dot Q_a}{k_B T_a}$.
In the case of infinite reservoirs, considered in \cite{Ja99}, this
leads to an ``apparent'' violation of the (FT) as pointed out in \cite{CV03a}
quantitatively in a special case and extended in \cite{BGGZ05} (see
the ``exponential tails in the latter references).
\\
(4) Therefore the Fluctuation Theorem holds for the physically
interesting $\sum_a\frac{\dot Q_a}{k_B T_a}$ {\it as a
  consequence} of its validity for $\s_{tot}$ and it becomes visible
over much shorter time scale, independent on the size of the
thermostats.
\\
(5) The above derivation of the (FT) is somewhat unsatisfactory
because one cannot turn it into a mathematically complete theorem even
if one is willing to make strong assumptions (like the assumption that
the system is ``Anosov'' in absence of external forces and of
thermostats): this is because the lack of a boundedness constraint
imposed on the kinetic energy in $\CC_0$ makes phase space unbounded; more
physically it is the above ``efficiency'' assumption on the interaction
between thermostats and system which is not properly understood
mathematically but which is necessary in the proof. 
In this respect the results here suffer from some of the 
drawbacks in the analysis in \cite{Ja99} (see comments after eq. (33)
and at the end of Sec. 4 of the latter reference) which also relies on
an efficiency assumption on the thermostats. Adding a constraint that
the total kinetic energy of the system in $\CC_0$ (as done in
\cite{Ga96a}) would trivially solve this problem but it would be
physically quite unsatisfactory.
\*

Coming back to the question of defining an irreversibility degree of a
 process $\G$ we distinguish between the (non stationary) state $\m_t$
 into which the initial state $\m_0$ evolves in time $t$, under
 varying forces and volume, and the state $\m_{SRB,t}$ obtained by
 ``freezing'' forces and volume at time $t$ and letting the system
 settle to become stationary, see coments following Eq. (\ref{e2.1}). 
We call $\e_t$ the entropy creation rate
 Eq. (\ref{e4.5}) and $\e^{srb}_t$ the entropy creation rate in the
 ``frozen'' state $\m_{SRB,t}$.

The proposal is to define {\it irreversibility degree} $\II(\G)$ or
{\it irreversibility time scale} $\II(\G)^{-1}$ of a process $\G$ by
setting:

\kern-3mm
\begin{eqnarray}\II(\G)=\frac1{N^2}\int_0^\infty
\Big(\media{\e_t}_{\m_t}-\media{\e^{srb}_{t}}_{SRB,t}\Big)^2
dt\label{e4.7}\end{eqnarray}
If the Chaotic Hypothesis is assumed then the state $\m_t$ will evolve
exponentially fast under the ``frozen evolution'' to
$\m_{SRB,t}$. Therefore the integral in Eq. (\ref{e4.7}) will converge for
reasonable $t$ dependences of $\BF,V$.

A physical definition of ``quasi static'' transformation is a
transformation that is ``very slow''. This can be translated
mathematically into an evolution in which $\BF_t$ evolves like, if not
exactly, as

\kern-3mm
\begin{eqnarray}\BF_t=\BF_0+ (1-e^{-\g t})
(\BF_\infty-\BF_0).\label{e4.8}\end{eqnarray}
An evolution $\G$ close to quasi static, but simpler for computing
$\II(\G)$, would proceed changing $\F_0$ into $\F_\infty=\F_0+\D$ by
$\D/\d$ steps of size $\d$, each of which has a time duration $t_\d$
long enough so that, at the $k$-th step, the evolving system settles
onto its stationary state at field $\F_0+k\d$. If the corresponding
time scale can be taken $=\k^{-1}$, independent of the value of the
field so that $t_\d$ can be defined by $\d e^{-\k t_\d}\ll 1$, then
$\II(\G)= const\, \d^{-1}\d^2\log\d^{-1}$ because the variation of
$\s_{(k+1)\d,+}-\s_{k\d,+}$ is, in general, of order $\d$ as a
consequence of the differentiability of the SRB states with respect to
the parameters, \cite{Ru97b}.  \*

\centerline{\bf 5, Comments and examples}

\0{\it Remarks:} {\bf(1)} A drawback of the definition proposed in
Sec.4 is that although $\media{\e^{srb}_t}_{SRB,t}$ is {\it
independent} on the metric that is used to measure volumes in phase
space the quantity $\media{\e_t}_{\m_t}$ {\it depends} on it. Hence the
irreversibility degree considered here reflects also properties of our
ability or method to measure (or to imagine to measure) distances in
phase space. One can keep in mind that a metric independent definition
can be simply obtained by minimizing over all possible choices of the
metric: but the above simpler definition seems, nevertheless, preferable.

\kern2mm \0{\bf(2)} It is $\II(\G)\ge0$, but I have not been able to
prove (except in the limit of a quasi static process) that $\II(\G)>0$:
this is a natural question to raise.

\kern2mm \0{\bf(3)} Suppose that a process takes place because of the
variation of an acting conservative force, for instance because a
gravitational force changes as a large mass is brought close to the
system, while no change in volume occurs and the thermostats have all
the same temperature. Then the "frozen" SRB distribution, for all $t$,
is given by Eq. (\ref{e3.2}) and $\media{\e^{srb}}_{SRB,t}=0$ (because the
"frozen equations" admit a SRB distribution which has a density, see
Eq. (\ref{e3.2}), in phase space). The isothermal process thus defined has {\it
therefore} $\II(\G)>0$ except, possibly, if it is {\it isentropic}, {\it i.e.}
if initial and final thermodynamic entropies are equal. In fact the
latter are cases in which the initial, intermediate and final states
are described by a density function $\r_t(x)$ on phase space hence,
\cite{Ru96}, it is $\media{\e^{srb}_t}_{SRB,t}=0$ and,\cite{An82},
${\media{\e_t}}_{\m_t}=-\frac{d}{dt}\int \r_t(x)\log\r_t(x) \,dx$. Since
the initial and final states are equilibrium states this means that if
the initial and final thermodynamic entropies ({\it i.e.} Gibbs') are
different it must be $\media{\e_t}_{\m_t}\ne0$ on a set of times of positive
measure and $\II(\G)>0$ and we expect it to be of order $\d$ (see last
comment in Sect.4). On the other hand if the thermodynamic
entropies are the same and $\G$ is almost quasi static then the
discussion at the end of Sec.4 shows that the irreversibility
time scale may tend to $+\infty$ when $\d\to0$ much faster than in the cases
in which the entropies differ ({\it e.g.} as $\d^{3}$ rather than as
$\d$).

\kern2mm \0{\bf(4)} The comment (3) shows that if a connection between
thermodynamic entropy and irreversibility scale can be established at
all it will neither be too direct nor too simple because in the
considered isothermal process linking two equilibrium states there is,
in general, a non zero entropy variation hence $\II(\G)>0$ but, if
performed very slowly, it can be close to be reversible both in the
classical Thermodynamics and in the irreversibility scale senses.

\kern2mm \0{\bf(5)} Consider a typical irreversible process. Imagine a
gas in an adiabatic cylinder covered by an adiabatic piston and
imagine to move the piston. The simplest situation arises if
gravity is neglected and the piston is suddenly moved at 
speed $w$.

Unlike the cases considered so far, the absence of thermostats
    (adiabaticity of the cylinder) imposes an extension of the
    analysis. In this case the phase space is the surface of constant
    energy in $\CC_0^{N_0}\times R^{3N_0}$ rather than the full
    $\CC_0^{N_0}\times R^{3N_0}$. Therefore if the piston moves at a
    speed which is too slow ({\it i.e.} slower than the maximum velocity of
    the particles on the energy surface) it will change the total
    energy of the gas in a way that depends on the special
    configuration initially chosen.


Therefore the simplest situation arises when the piston is moved at
speed so large that no energy is gained or lost by the particles
because of the collisions with the moving wall (this is, in fact, a
> case in which there are no such collisions). This is an extreme 
idealization of the  
classic Joule-Thomson experiment.

Let $S$ be the section of the cylinder and $H_t = H_0+w\,t$ be the
distance between the moving lid and the opposite base. Let $\Omega =
S\,H_t$ be the cylinder volume. In this case the volume of phase space
changes only because the boundary moves and it increases by $N \,w\,S\,\Omega^{
N-1}$ per unit time, {\it i.e.} its rate of increase is
$N \frac{w}{H_t}$.

Hence $\media{\e_t}_t$ is $-N \frac{w}{H_t}$, while $\e^{srb}_t\equiv 0$. If
$T=\frac{L}w$ is the duration of the transformation ("Joule-Thomson" pro-cess)
increasing the cylinder length by $L$, then

\kern-5mm
\begin{eqnarray}\txt
\II(\Gamma ) =N^{-2}\int_0^T N^2\big(\frac{w}{H_t}\big)^2\,dt 
\tende{T\to\infty} 
w\frac{L}{H_0(H_0+L)}\label{e5.1}\end{eqnarray}
\0and the transformation is irreversible. The irreversibility time scale
approaches $0$ as $w\to\infty$, as possibly expected.  If $H_0 = L$,
i.e. if the volume of the container is doubled, then $I(\G) =
\frac{w}{2L}$ and the irreversibility time scale of the process coincides
with its "duration". 

\0(6) A different situations arises if in the context of (5) the
piston is replaced by a sliding lid which divides the cylinder in two
halves of height $L$ each: one empty at time zero and the other
containing the gas in equilibrium. At time $0$ the lid is lifted and a
process $ \Gamma'$ takes place. In this case $\dot V_t = V \d(t)$
because the volume $V = S\,L$ becomes suddenly double.  Therefore the
evaluation of the irreversibility scale yields

\kern-3mm
\begin{eqnarray}\II(\Gamma') = N^{ -2}\int_0^\infty N^2\d(t)^2\,dt
\equiv +\infty\label{e5.2}
\end{eqnarray}
so that the irreversibility becomes immediately manifest, $I(\Gamma')
= +\infty, \,\II(\G')^{-1}=0$. This idealized experiment is rather close
to the actual Joule-Thomson experiment.

In the latter example it is customary to estimate the degree of
irreversibility at the lift of the lid by the {\it thermodynamic
equilibrium entropy} variation between initial and final states. It
would of course be interesting to have a general definition of entropy
of a non stationary state (like the states $\m_t$ at times
$(t\in(0,\infty)$ in the example just discussed) that would allow
connecting the degree of irreversibility to the thermodynamic entropy
variation in processes leading from an initial equilibrium state to a
final equilibrium state, see \cite{GL03}.

The time scale introduced in Eq.(\ref{e4.7}) refers to the entire
process rather than to what happens to the system only; it might have
alternative interpretations: but it seems a quantity of interest in
itself.

\*
\0{\it Acknowledgements: I am indebted to F. Bonetto, A. Giuliani and
  F. Zamponi for many ideas, discussions and for earlier collaborations that
  preceded and stimulated the development of this work and to a
  referee for pointing out the reference to Jarzynski's paper, \cite{Ja99}.} 


\begin{thebibliography}{20}
\expandafter\ifx\csname natexlab\endcsname\relax\def\natexlab#1{#1}\fi
\expandafter\ifx\csname bibnamefont\endcsname\relax
  \def\bibnamefont#1{#1}\fi
\expandafter\ifx\csname bibfnamefont\endcsname\relax
  \def\bibfnamefont#1{#1}\fi
\expandafter\ifx\csname citenamefont\endcsname\relax
  \def\citenamefont#1{#1}\fi
\expandafter\ifx\csname url\endcsname\relax
  \def\url#1{\texttt{#1}}\fi
\expandafter\ifx\csname urlprefix\endcsname\relax\def\urlprefix{URL }\fi
\providecommand{\bibinfo}[2]{#2}
\providecommand{\eprint}[2][]{\url{#2}}

\bibitem[{\citenamefont{Jarzynski}(1999)}]{Ja99}
\bibinfo{author}{\bibfnamefont{C.}~\bibnamefont{Jarzynski}},
  \bibinfo{journal}{Journal of Statistical Physics}
  \textbf{\bibinfo{volume}{98}}, \bibinfo{pages}{77} (\bibinfo{year}{1999}).

\bibitem[{\citenamefont{Gallavotti}(1998)}]{Ga98}
\bibinfo{author}{\bibfnamefont{G.}~\bibnamefont{Gallavotti}},
  \bibinfo{journal}{Physica D} \textbf{\bibinfo{volume}{112}},
  \bibinfo{pages}{250} (\bibinfo{year}{1998}).

\bibitem[{\citenamefont{Gallavotti}(2004)}]{Ga04}
\bibinfo{author}{\bibfnamefont{G.}~\bibnamefont{Gallavotti}},
  \bibinfo{journal}{cond-mat/0402676}  (\bibinfo{year}{2004}).

\bibitem[{\citenamefont{Ruelle}(1999)}]{Ru99}
\bibinfo{author}{\bibfnamefont{D.}~\bibnamefont{Ruelle}},
  \bibinfo{journal}{Journal of Statistical Physics}
  \textbf{\bibinfo{volume}{95}}, \bibinfo{pages}{393} (\bibinfo{year}{1999}).

\bibitem[{\citenamefont{Evans and Morriss}(1990)}]{EM90}
\bibinfo{author}{\bibfnamefont{D.}~\bibnamefont{Evans}} \bibnamefont{and}
  \bibinfo{author}{\bibfnamefont{G.}~\bibnamefont{Morriss}},
  \emph{\bibinfo{title}{Statistical Mechanics of Non{\-}equilibrium Fluids}}
  (\bibinfo{publisher}{Academic Press}, \bibinfo{address}{New-York},
  \bibinfo{year}{1990}).

\bibitem[{\citenamefont{Gallavotti}(2005)}]{Ga05}
\bibinfo{author}{\bibfnamefont{G.}~\bibnamefont{Gallavotti}},
  \bibinfo{journal}{cond-mat/0510027}  (\bibinfo{year}{2005}).

\bibitem[{\citenamefont{G.Gallavotti}(1999)}]{Ga99}
\bibinfo{author}{\bibnamefont{G.Gallavotti}}, \bibinfo{journal}{Open Systems
  and Information Dynamics} \textbf{\bibinfo{volume}{6}}, \bibinfo{pages}{101}
  (\bibinfo{year}{1999}).

\bibitem[{\citenamefont{J.P.Eckmann et~al.}(1999)\citenamefont{J.P.Eckmann,
  Pillet, and Bellet}}]{EPR99}
\bibinfo{author}{\bibnamefont{J.P.Eckmann}},
  \bibinfo{author}{\bibfnamefont{C.}~\bibnamefont{Pillet}}, \bibnamefont{and}
  \bibinfo{author}{\bibfnamefont{L.~R.} \bibnamefont{Bellet}},
  \bibinfo{journal}{Communications in Mathematical Physics}
  \textbf{\bibinfo{volume}{201}}, \bibinfo{pages}{657} (\bibinfo{year}{1999}).

\bibitem[{\citenamefont{Gallavotti and Cohen}(2004)}]{GC04}
\bibinfo{author}{\bibfnamefont{G.}~\bibnamefont{Gallavotti}} \bibnamefont{and}
  \bibinfo{author}{\bibfnamefont{E.}~\bibnamefont{Cohen}},
  \bibinfo{journal}{Physical Review E} \textbf{\bibinfo{volume}{69}},
  \bibinfo{pages}{035104 (+4)} (\bibinfo{year}{2004}).

\bibitem[{\citenamefont{Bonetto et~al.}(2005)\citenamefont{Bonetto, Gallavotti,
  Giuliani, and Zamponi}}]{BGGZ05}
\bibinfo{author}{\bibfnamefont{F.}~\bibnamefont{Bonetto}},
  \bibinfo{author}{\bibfnamefont{G.}~\bibnamefont{Gallavotti}},
  \bibinfo{author}{\bibfnamefont{A.}~\bibnamefont{Giuliani}}, \bibnamefont{and}
  \bibinfo{author}{\bibfnamefont{F.}~\bibnamefont{Zamponi}},
  \bibinfo{journal}{cond-mat/0507672}  (\bibinfo{year}{2005}).

\bibitem[{\citenamefont{Gallavotti and Cohen}(1995)}]{GC95}
\bibinfo{author}{\bibfnamefont{G.}~\bibnamefont{Gallavotti}} \bibnamefont{and}
  \bibinfo{author}{\bibfnamefont{E.}~\bibnamefont{Cohen}},
  \bibinfo{journal}{Physical Review Letters} \textbf{\bibinfo{volume}{74}},
  \bibinfo{pages}{2694} (\bibinfo{year}{1995}).

\bibitem[{\citenamefont{Gallavotti et~al.}(2004)\citenamefont{Gallavotti,
  Bonetto, and Gentile}}]{GBG04}
\bibinfo{author}{\bibfnamefont{G.}~\bibnamefont{Gallavotti}},
  \bibinfo{author}{\bibfnamefont{F.}~\bibnamefont{Bonetto}}, \bibnamefont{and}
  \bibinfo{author}{\bibfnamefont{G.}~\bibnamefont{Gentile}},
  \emph{\bibinfo{title}{Aspects of the ergodic, qualitative and statistical
  theory of motion}} (\bibinfo{publisher}{Springer Verlag},
  \bibinfo{address}{Berlin}, \bibinfo{year}{2004}).

\bibitem[{\citenamefont{Zon and Cohen}(2003)}]{CV03a}
\bibinfo{author}{\bibfnamefont{R.~V.} \bibnamefont{Zon}} \bibnamefont{and}
  \bibinfo{author}{\bibfnamefont{E.}~\bibnamefont{Cohen}},
  \bibinfo{journal}{Physical Review Letters} \textbf{\bibinfo{volume}{91}},
  \bibinfo{pages}{110601 (+4)} (\bibinfo{year}{2003}).

\bibitem[{\citenamefont{Ruelle}(1996)}]{Ru96}
\bibinfo{author}{\bibfnamefont{D.}~\bibnamefont{Ruelle}},
  \bibinfo{journal}{Journal of Statistical Physics}
  \textbf{\bibinfo{volume}{85}}, \bibinfo{pages}{1} (\bibinfo{year}{1996}).

\bibitem[{\citenamefont{Ruelle}(1997{\natexlab{a}})}]{Ru97}
\bibinfo{author}{\bibfnamefont{D.}~\bibnamefont{Ruelle}},
  \bibinfo{journal}{Communications in Mathematical Physics}
  \textbf{\bibinfo{volume}{189}}, \bibinfo{pages}{365}
  (\bibinfo{year}{1997}{\natexlab{a}}).

\bibitem[{\citenamefont{Gallavotti}(1995)}]{Ga95b}
\bibinfo{author}{\bibfnamefont{G.}~\bibnamefont{Gallavotti}},
  \bibinfo{journal}{Mathematical Physics Electronic Journal (MPEJ)}
  \textbf{\bibinfo{volume}{1}}, \bibinfo{pages}{1} (\bibinfo{year}{1995}).

\bibitem[{\citenamefont{Gallavotti}(1996)}]{Ga96a}
\bibinfo{author}{\bibfnamefont{G.}~\bibnamefont{Gallavotti}},
  \bibinfo{journal}{Physical Review Letters} \textbf{\bibinfo{volume}{77}},
  \bibinfo{pages}{4334} (\bibinfo{year}{1996}).

\bibitem[{\citenamefont{Ruelle}(1997{\natexlab{b}})}]{Ru97b}
\bibinfo{author}{\bibfnamefont{D.}~\bibnamefont{Ruelle}},
  \bibinfo{journal}{Communications in Mathematical Physics}
  \textbf{\bibinfo{volume}{187}}, \bibinfo{pages}{227}
  (\bibinfo{year}{1997}{\natexlab{b}}).

\bibitem[{\citenamefont{Andrej}(1982)}]{An82}
\bibinfo{author}{\bibfnamefont{L.}~\bibnamefont{Andrej}},
  \bibinfo{journal}{Physics Letters} \textbf{\bibinfo{volume}{111A}},
  \bibinfo{pages}{45} (\bibinfo{year}{1982}).

\bibitem[{\citenamefont{Goldstein and Lebowitz}(2004)}]{GL03}
\bibinfo{author}{\bibfnamefont{S.}~\bibnamefont{Goldstein}} \bibnamefont{and}
  \bibinfo{author}{\bibfnamefont{J.}~\bibnamefont{Lebowitz}},
  \bibinfo{journal}{Physica D} \textbf{\bibinfo{volume}{193}},
  \bibinfo{pages}{53} (\bibinfo{year}{2004}).

\end{thebibliography}
\bibliographystyle{apsrev} 

\revtex

\end{document}